# Defect Detection Network In PCB Circuit Devices Based on GAN Enhanced YOLOv11

Jiayi Huang[1,4], Feiyun Zhao[2,5], Lieyang Chen[3,6]

[1]College of Electronic Information Engineering, Anhui University, China
[2]George Washington University, Washington, DC, USA
[3]Columbia University in the City of New York, New York, USA

[4]hjyyyy_93@163.com
[5]feiyun.zhao@gwmail.gwu.edu
[6]lc3548@columbia.edu

**Abstract.** This study proposes an advanced method for surface defect detection in printed circuit boards (PCBs) using an improved YOLOv11 model enhanced with a generative adversarial network (GAN). The approach focuses on identifying six common defect types: missing hole, rat bite, open circuit, short circuit, burr, and virtual welding. By employing GAN to generate synthetic defect images, the dataset is augmented with diverse and realistic patterns, improving the model's ability to generalize, particularly for complex and infrequent defects like burrs. The enhanced YOLOv11 model is evaluated on a PCB defect dataset, demonstrating significant improvements in accuracy, recall, and robustness, especially when dealing with defects in complex environments or small targets. This research contributes to the broader field of electronic design automation (EDA), where efficient defect detection is a crucial step in ensuring high-quality PCB manufacturing. By integrating advanced deep learning techniques, this approach enhances the automation and precision of defect detection, reducing reliance on manual inspection and accelerating design-to-production workflows. The findings underscore the importance of incorporating GAN-based data augmentation and optimized detection architectures in EDA processes, providing valuable insights for improving reliability and efficiency in PCB defect detection within industrial applications.

## 1.Introduction

The quality of printed circuit boards (PCBs) is pivotal to the reliability and stability of electronic devices, as defects in PCBs can lead to electrical malfunctions, ultimately compromising the performance of the associated equipment. To address the growing demand for higher quality standards in PCB manufacturing, traditional manual inspection methods are increasingly being replaced by automated intelligent inspection technologies. In the domain of Electronic Design Automation (EDA), PCB defect detection plays a vital role in ensuring the quality of both design and production processes. Advanced inspection techniques significantly enhance production efficiency, reduce human error, and streamline workflows from design to manufacturing.

In recent years, deep learning has emerged as a transformative technology for defect detection, with significant success in tasks such as image classification and object detection. The YOLO (You Only Look Once) algorithm, renowned for its high efficiency and real-time detection capabilities, has gained prominence in PCB defect detection. YOLO simplifies object detection by converting it into a single-stage regression task, achieving a remarkable balance between speed and accuracy. This makes it highly effective for PCB scenarios involving challenges such as intricate backgrounds, small-scale defects, and coexisting multiple defect types. Furthermore, the evolution of YOLO into advanced versions like YOLOv4, YOLOv5, and YOLOv11 has enhanced its ability to extract features and generalize to complex patterns, making it an indispensable tool in PCB defect detection.

This study builds on the PCB defect dataset provided by the Peking University Open Laboratory of Human-Computer Interaction to propose a novel defect detection method. By integrating a generative adversarial network (GAN) into the YOLOv11 framework, the model benefits from the GAN's ability to generate synthetic defect images, effectively augmenting the dataset. This augmentation enhances the model's capacity to learn diverse and intricate defect patterns, particularly excelling in detecting rare and complex defects such as burrs and virtual welds.

In the context of EDA, the proposed method not only improves the precision and robustness of PCB defect detection but also supports the automation and intelligentization of the PCB design-to-manufacturing process. This research demonstrates the efficacy of GAN-enhanced YOLOv11 and provides valuable insights for advancing EDA technologies in industrial applications.

## 2. Literature Review

In recent years, the surface defect detection technology of printed circuit board (PCB) has made remarkable progress, especially the target detection method based on deep learning, which greatly improves the accuracy and efficiency of detection. In this process, the improvement and application of YOLO(You Only Look Once) series models have been widely concerned. In this paper, combined with the existing research, the application of YOLO model and its improved method in PCB defect detection is summarized in detail.

In recent years, Tang et al. proposes PCB-YOLO, an improved YOLOv5-based algorithm for PCB defect detection, incorporating K-means++ anchors, small target detection layers, Swin transformer, and EIoU loss to enhance accuracy, speed, and model efficiency, achieving 95.97% mAP at 92.5 FPS for real-time, high-precision defect detection [1].Wei Chen, Zhongtian Huang, Mu Qian, and Yi Sun proposed a Transformer-YOLO-based PCB defect detection model that utilizes an improved clustering algorithm for anchor box generation, Swin Transformer for feature extraction, and a convolution-attention mechanism module, resulting in significantly improved detection accuracy [2].

Shiyi Luo and colleagues proposed an enhanced YOLOv7 network (EC-YOLOv7) for detecting and classifying electronic components (ECs) on PCBs. By integrating ACmix modules into the E-ELAN architecture, improving SPPCSPS with concurrent channels, employing the DyHead for spatial information capture, and introducing the WIoU-Soft-NMS regression method, they improved detection accuracy and speed. The EC-YOLOv7 network achieved a mAP@0.5 of 94.4% and higher FPS on the PCB dataset, significantly surpassing the original YOLOv7 and other EC detection methods [3].

Minghao Yuan et al. proposed a novel YOLO-HMC network based on an improved YOLOv5 framework to address the challenge of accurately detecting tiny PCB defects in complex backgrounds. By incorporating HorNet for feature extraction, MCBAM to highlight defect locations, and CARAFE for improved contextual semantic aggregation, they optimized the detection head to enhance precision. The YOLO-HMC achieved a mAP of 98.6% on public PCB defect datasets, surpassing state-of-the-art models in accuracy and efficiency [4].

Bowei Du et al. proposed an enhanced YOLOv5s-based network, YOLO-MBBi, to improve PCB surface defect detection accuracy and real-time performance. By integrating MBConv modules, CBAM attention, BiFPN, depth-wise convolutions, and replacing CIoU with SIoU loss, YOLO-MBBi achieved mAP50 and recall values of 95.3% and 94.6%, respectively—outperforming YOLOv5s by 3.6% and 2.6%. With significantly reduced FLOPs (12.8) and a high FPS value of 48.9, the model met industrial production requirements [5].

In the field of combining deep learning with PCB defect detection, A analyzed the application potential of the latest YOLO model in PCB defect detection [6]. Kewen Xia and colleagues proposed GCC-YOLO, a YOLO-based model enhanced with global contextual attention and ConvMixer prediction heads to improve PCB tiny target detection in complex backgrounds. By incorporating a high-resolution P2 layer, GC module, BiFPN feature fusion, and ConvMixer-based prediction heads, GCC-YOLO enhanced small target detection while reducing parameters. On the PCB dataset, it improved Precision, Recall, mAP@0.5, and mAP@0.5:0.95 by 0.2%, 1.8%, 0.5%, and 8.3%, respectively, compared to YOLOv5s, achieving better performance and faster inference speed [7].

## 3. Data and Model

*3.1 Dataset and Preprocessing*
This study utilizes the PCB defect dataset from Peking University, containing 1,386 labeled images across six defect types: missing hole, mouse bite, open circuit, short circuit, burr, and virtual welding. To improve model generalization, the dataset is augmented using a GAN, which generates realistic defect images with diverse shapes, sizes, and positions. The GAN's generator simulates defects based on real PCB features, while the discriminator refines the quality. This augmentation expands defect diversity, particularly for complex defects like burrs, while preserving the original distribution. The enhanced dataset improves the model's learning, boosting detection accuracy and robustness.

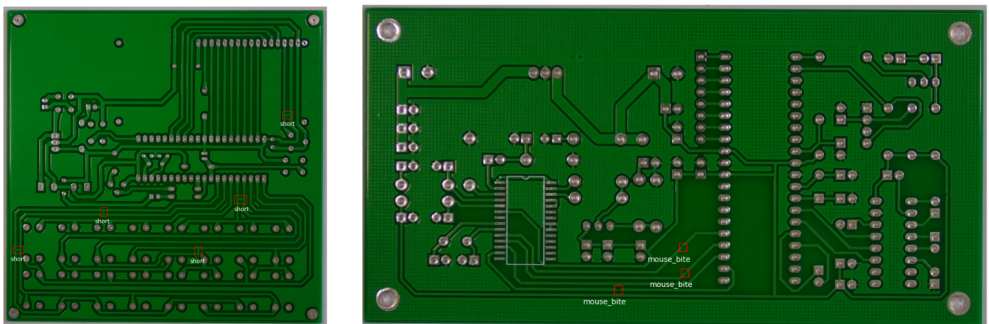

**Figure 1.** Dataset Display

*3.2 Improved YOLOv11 Model for PCB Defect Detection*
This study utilizes the YOLOv11 model for PCB defect detection. As the latest version of the YOLO series, YOLOv11 maintains its fast, end-to-end detection characteristics while introducing several improvements to balance speed and accuracy. Its "one-shot" design allows predictions for multiple targets and categories through a single forward pass of the network, making it highly efficient and well-suited for detecting complex PCB defects.YOLOv11 employs a dynamic anchor box allocation strategy, dynamically assigning anchor boxes during training to adaptively learn target shapes and sizes. This strategy addresses anchor mismatch issues and improves detection accuracy, particularly for complex defects such as burrs and virtual welding. This capability enhances the model's performance in recognizing small and intricate targets on PCBs [8].

The model features a new backbone network combining the strengths of lightweight networks (e.g., MobileNet) and deep convolutional networks (e.g., ResNet). This hybrid architecture balances computational efficiency with enhanced feature extraction, enabling precise detection of small defects and distinguishing them from background noise. YOLOv11 adapts well to diverse PCB defect types, ensuring flexibility across different scenarios.YOLOv11 also optimizes multi-scale prediction, leveraging detection at various feature layers to improve performance on targets of different sizes. This capability ensures robust detection across complex circuit layouts, addressing both large-scale defects like short circuits and small-scale ones like mouse bites.Finally, YOLOv11 introduces a loss function combining Focal Loss and IoU Loss, improving robustness against class imbalance and enhancing bounding box prediction accuracy. This loss function effectively reduces background noise interference, further boosting detection reliability.

These advancements make YOLOv11 highly effective in PCB defect detection, improving accuracy, recall, and robustness. The model supports automated quality control and production, offering a reliable solution for detecting diverse PCB defects.

*3.3 Method Steps*

To detect PCB defects, this study adopts YOLOv11 as the foundational model, specifically optimized for the task of PCB defect detection. YOLOv11, known for its high detection accuracy and fast processing speed, has undergone customization and enhancements to better suit the requirements of PCB defect detection. The training and evaluation process consists of the following key steps:

Data Preprocessing: The augmented dataset (comprising the original dataset and GAN-generated defect images) was normalized and enhanced with transformations such as rotation, scaling, and contrast adjustment. By integrating simulated defect images generated by GAN, the model's robustness to diverse PCB image variations was improved. The dataset was randomly divided into training and validation sets to ensure the model's ability to generalize to unseen samples.

Model Architecture Modifications: Custom anchor boxes tailored to the size distribution of PCB defects (including GAN-generated defects) were introduced to enhance YOLOv11's detection capabilities. The feature extraction layers were modified by adding convolutional layers and adjusting kernel parameters to better capture fine-grained PCB defect features. These modifications were particularly effective for detecting diverse and complex defects such as those simulated by GAN. Additionally, activation functions and layer configurations were refined to increase sensitivity to subtle anomalies.

Model Training: The model was trained using an innovative optimizer, Nadam (Nesterov-accelerated Adaptive Moment Estimation). Nadam combines the strengths of Adam and Nesterov-accelerated gradient, providing faster convergence and better performance, especially in complex backgrounds and small-target detection tasks [9].A dynamic learning rate mechanism was employed, with an initial learning rate of 0.001 adjusted using the Cosine Annealing strategy. This approach starts with a higher learning rate for global exploration, gradually reducing it to fine-tune local optimization. This strategy improves convergence speed and avoids local minima.During training, three loss functions were monitored: box loss, class loss, and distribution focal loss. The distribution focal loss was adjusted to focus more on learning complex defect samples, particularly from GAN-generated data, effectively balancing defect distribution.

Post-Processing and Threshold Adjustment: Non-Maximum Suppression (NMS) thresholds were fine-tuned to minimize false positives. An adaptive confidence thresholding mechanism was implemented, dynamically adjusting thresholds based on the characteristics of different defect categories, including GAN-generated defects, for better distinction between defect and non-defect areas.

Model Evaluation: The model was evaluated using metrics such as mean Average Precision (mAP), precision, and recall. mAP was calculated over thresholds from 0.5 to 0.95 to comprehensively assess detection accuracy across varying levels of overlap, with special attention to the improvement in detecting complex GAN-simulated defects.

By integrating an advanced optimizer and dynamic learning rate mechanism, YOLOv11 achieved significant performance improvements in PCB defect detection. The model demonstrated enhanced accuracy and robustness in handling complex defect scenarios and diverse defect distributions, providing strong support for automated quality control and PCB production workflows.

## 4. MODEL RESULT ANALYSIS

Figure 2 illustrates the changes in Box Loss and Class Loss for the training and validation sets. Box Loss measures the accuracy of bounding box predictions in locating defects. As training progresses, the training set loss decreases sharply between the 0th and 25th epoch, indicating rapid learning of preliminary features. It stabilizes around 1.0, suggesting the model is maturing in positioning. However, validation set loss fluctuates significantly, especially after the 50th epoch (1.5–2.0), indicating limited generalization and potential overfitting on some samples.

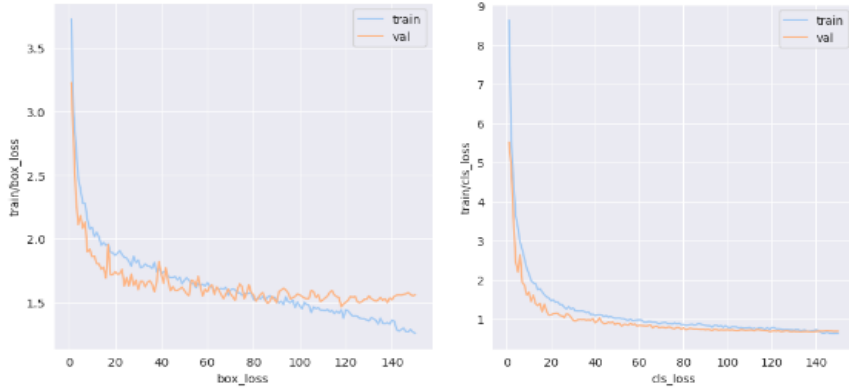

**Figure 2.** Box Loss and Class Loss

Figure 2 shows the change in Class Loss. The training set loss quickly drops near zero, indicating the model efficiently learns defect classification early. The validation set loss fluctuates initially but stabilizes in later epochs, suggesting some early variability in category prediction. Overall, the model effectively distinguishes defect types, demonstrating its ability to classify defects accurately across the dataset despite initial fluctuations in validation performance.

**Table 1.** Model index analysis

| Class | Images | Instances | P | R | mAP50 | mAP50-95 |
|---|---|---|---|---|---|---|
| all | 66 | 278 | 0.95 | 0.87 | 0.92 | 0.49 |
| missing_hole | 11 | 46 | 0.83 | 0.64 | 0.75 | 0.39 |
| mouse_bite | 11 | 48 | 1.00 | 0.88 | 0.94 | 0.50 |
| open_circuit | 13 | 55 | 0.94 | 0.85 | 0.94 | 0.45 |
| short | 11 | 45 | 0.99 | 0.98 | 0.99 | 0.55 |
| spur | 9 | 37 | 0.99 | 1.00 | 0.99 | 0.50 |
| spurious_copper | 11 | 47 | 0.94 | 0.87 | 0.94 | 0.40 |

Missing hole: the detection accuracy of missing hole category is the lowest, P=0.83, R=0.64, mAP50 is 0.75, and mAP50-0.95 is 0.39. This indicates that the performance of the model in detecting and recognizing hole defects needs to be improved

Mouse bite: the detection accuracy of rat bite category is the highest, P=1.00, R=0.887, mAP50 is 0.94, and mAP50-0.95 is 0.50. Although the overall performance is good, compared with the lack of holes, the performance of rat bite decreases at a higher threshold (mAP_50:0.95), which indicates that the model may not be able to fully capture the small or blurred rat bite defects in details.

Open circuit: the detection performance of open circuit is relative high, P=0.94, R=0.85, mAP50 is 0.94, and mAP50-0.95 is 0.445. The detection accuracy is high, but in more complex samples, the model may not be able to deal with open defects, especially those with unclear boundaries.

Short: the detection performance of short-circuit defects is ideal, with P=0.99, R=0.98, mAP50 = 0.99 and mAP50-0.95 = 0.55. This shows that the model shows high stability in dealing with short-circuit defects and can accurately identify and classify short-circuit defects.

Spur: the detection performance of burr category is still high, with P=0.99, R=1.00, mAP50 being 0.99, and mAP50-0.95 being 0.50. This indicates that although the features of burrs are complex and irregular in shape, the model can still accurately classify them.

Spurious copper: The detection performance of virtual welding is quite satisfactory, with P=0.94, R=0.87, mAP50 = 0.94 and mAP50-0.95 = 0.40. It shows that the model has certain robustness in dealing with this kind of obvious defect.

## 5.Conclusions

This study explores PCB defect detection using a GAN-enhanced YOLOv11 model, achieving notable success. By leveraging GAN to generate diverse defect data, the model effectively handles complex defect types like burrs, significantly improving generalization. Across six defect categories (missing hole, mouse bite, open circuit, short circuit, burr, and spurious copper), it demonstrates strong performance with an average precision (P) of 0.95 and recall (R) of 0.87, reflecting high accuracy and stability. Analysis of loss curves highlights discrepancies between training and validation sets, indicating areas for improvement in generalization. Future enhancements could include more diverse training data, advanced augmentation techniques, and parameter optimization.

As a deep learning-based object detection algorithm, YOLOv11 exhibits substantial potential in PCB defect detection, excelling in identifying multiple defects within complex backgrounds and small targets. The integration of GAN-generated data enhances its ability to capture fine-grained features, making it a valuable tool in electronic design automation (EDA). In this context, intelligent PCB defect detection boosts circuit design reliability, reduces manufacturing losses, and accelerates design iterations. This study provides an efficient detection solution for EDA, advancing intelligent electronic design and automated technologies. The GAN-enhanced YOLOv11 model serves as a robust reference for optimizing EDA workflows, with future work focusing on improving performance for diverse defects and integration with other EDA tools.